\documentclass[12pt,preprint]{aastex}
\usepackage{graphicx}
\usepackage{longtable}

\begin{document}

\newcommand{\zdot}{\makebox[0pt][l]{.}}
\newcommand{\up}[1]{\ifmmode^{\rm #1}\else$^{\rm #1}$\fi}
\newcommand{\dn}[1]{\ifmmode_{\rm #1}\else$_{\rm #1}$\fi}
\newcommand{\upd}{\up{d}}
\newcommand{\uph}{\up{h}}
\newcommand{\upm}{\up{m}}
\newcommand{\ups}{\up{s}}
\newcommand{\arcd}{\ifmmode^{\circ}\else$^{\circ}$\fi}
\newcommand{\arcm}{\ifmmode{'}\else$'$\fi}
\newcommand{\arcs}{\ifmmode{''}\else$''$\fi}

\pagestyle{plain}

\def\thefootnote{\fnsymbol{footnote}}

\title{    Search for QSO candidates in OGLE-II data.
                            }
               \author{Laurent Eyer}
\affil{Princeton University Observatory, Princeton, NJ 08544, USA}

\begin{abstract}
A search for faint slowly variable objects was undertaken in the hope
of finding QSO candidates behind the Small and Large Magellanic Clouds
(SMC and LMC). This search used the optical variability properties of
point sources from the Magellanic cloud OGLE-II photometric
data. Objects bluer than $V-I=0.9$ and within $ 17 < I < 20.5 $ were
studied. Robust variograms/structure functions have been computed for
each time-series and only candidates showing a significant increasing
variability over longer time scales were selected.

Several light curves were identified as having probable artifacts and
were therefore removed.  Stars showing signs of periodicity or small
trends in their light curves were also removed and we are left with
mostly either Be stars ($\gamma\,$Cas stars) or QSO candidates.

We present a list of 25 slowly varying objects for SMC and 155 for
LMC, out of 15'000 and 53'000 variable objects respectively. Of these,
about 15 objects for the SMC and 118 objects for the LMC are QSO candidates.
\end{abstract}

\section{Introduction}

The main motivation of this study is to find QSOs behind the
Magellanic clouds. Few QSOs behind SMC and LMC are known.  Several
previous studies have reported such discoveries with their associated
coordinates, but none of them was in OGLE fields. However more
recently, Dobrzycki et al. (2002) found four new QSOs thanks to
spectroscopic observations of the optical counterpart of X-ray
sources. Those four candidates were observed by OGLE-II.

Other studies have been undertaken for several years by the MACHO
group (Geha et al. 1999, Drake et al. 2001). They found about 30 QSOs
behind LMC, but have not yet published their coordinates.

QSOs can be useful for different purposes, for example: for fixing
the coordinate system in proper motion studies of LMC, SMC or
foreground stars; for mapping the interstellar material in the
Magellanic Clouds.

The optical variability properties of QSOs are not
very well known. Some efforts are underway to remedy this situation;
for instance AGNs are monitored in 11 passbands by Kobayashi et
al. 1998 (MAGNUM project).

Most QSOs are variable (Cristiani et al. 1996). Discovery of QSOs have
been achieved using their optical variability (Trevese et al. 1989,
Brunzendorf \& Meusinger 2001).  We note however that only one QSO out
of the four in Dobrzycki's studies was classified as a variable in
OGLE-II. This is mainly due to the faintness of the three other
objects (about $V \simeq 20$) and the resulting large errors ($\simeq
0.1$) in the OGLE-II photometry.

We exploit the variability properties of QSOs since photometric
data of OGLE-II is in public domain and offers a 4 years observing
period with about 300-400 data points per object in $I$-band and about
30 in $B$ and $V$-bands.

From SDSS, we know (Strauss 2002) that there are about 5 QSOs per
square degree brighter than $i = 19$ mag. The number is smaller by a
factor 5 per magnitude. As SDSS $i$-mag is approximately $I$-mag, a
cut $I > 17$ will exclude one or two QSOs in all fields together,
and therefore seems a sensible limit. We should expect to find about
30 QSOs behind the SMC and LMC brighter than $I = 19$ mag.

\section{Selection criteria}

A schematic description of the different selection criteria can be
found in Fig.~1. The criteria are described in the
following sections.

\subsection{Selection on photometry}
\label{sec:magsel}
The OGLE-II experiment measured a total of 9 million stars in the
direction of the Magellanic clouds. The LMC catalog lists 7 million
star coordinates covering 5.7 square deg and B, V, I photometry
(Udalski et al. 2000), and SMC data catalog lists 2.2 million star
coordinates covering 2.4 square deg (Udalski et al. 1998). The
colour-magnitude diagrams for the stars observed by OGLE-II (subsample
selected randomly) and for the variable stars (see
Fig.~2) have different distributions. The variable star
catalog lists 68'000 stars: 53'000 for LMC and 15'000 for SMC,
(cf. \.Zebru\'n et al. 2001). The variability analysis covered 21
fields (4.5 square deg) out of 26 fields for LMC data (the LMC fields
22 to 26 were monitored only on 13 nights). The pipeline for the
OGLE-II analysis using Difference Image Analysis and the criterion for
selecting the variable stars are described by Wo\'zniak (2000). The
two lower diagrams of Fig.~2 (left LMC and right SMC)
are quite stunning. Though we recognize variables from the red giant
branch (right border of the diagram), RR Lyrae stars (spherical clump
at $I = 18.7$ and $V-I =0.55$ for LMC, mostly too faint in SMC),
Cepheids (the rather vertical clump/strip above RR Lyrae stars), the B
variable stars (left border), we also notice that the main sequence
seems to separate in two branches, that a population of very bright
variable objects forms a strip (to the right of Cepheid strip), and
that the red giants are quite clumped (we further point out that the
red giant tip produces a discontinuity between the giant branch and
the asymptotic giant branch). Those aspects will be studied in a
separate article.  For our purpose, we want to explore the lower part
($I > 17$) of the colour-magnitude diagrams (see dashed lines of
Fig.~2). The magnitude cut selects, for the LMC data, a
population of main sequence blue variables, RR Lyrae stars and red
giant stars. Some short period Cepheid are included in the selected
sample of SMC data because it is further away and has a lower
metallicity.

In order to reject the time series of the red giant branch variable
stars which may also have slow variations as QSOs do, we select
objects bluer than $V-I=0.9$ and we also select the objects brighter
than $I=20.5$.  To avoid any misidentification problem, all stars
which have a position further than 1 pixel ($0.417$ arcsec) from the
other determinations of position (matches of dophot and DIA position)
were removed.  We show in Fig.~3 the histogram of
those distances.  With these criteria, we get a first selection of
6241 objects for LMC and 1553 objects for SMC out of the samples of
53'000 (LMC) and 15'000 (SMC) variable objects.

\subsection{Selection on time variations}
\label{sec:timescale}

It is known that generally QSOs have little variability at short time
scales, and that the variability, though irregular and aperiodic, is
increasing when longer time scales are observed. We note however that
the BL Lac objects could have a variability of several tenth of a
magnitude on a day to day basis.

The criterion of selection is an increase of variability for time
scales longer than 100 days and the employed mathematical tool is the
variograms/structure functions. This permits us to determine the time
scale of the variations in a given signal (Hughes et al. 1992, Eyer \&
Genton 1999).  All pair differences of time, $h_{ij}=JD_j-JD_i$ ($h$
is the lag; JD is the Julian Date) and squared pair differences of
magnitude, $(I_j-I_i)^2$, are computed.

Given a lag $h$, we compute the median of the subsample,
$2\gamma_{\mbox{\tiny med}}(h)$, of $(I_j-I_i)^2$ formed by all
possible pairs $i, j$ such that $h_{ij} < h$. It gives an estimation
of the spread of the distribution formed by the $I_j-I_i$.  We call
this function variogram.  If the distribution of the $I_j-I_i$ is
symmetric then the square root of $2 \gamma_{\mbox{\tiny med}}(h)$ is
the interquartile range of the distribution of the $I_j-I_i$.  For an
example see Fig.~4. The slope of the variogram can
be used as a discriminating criterion for selecting the time series where
variability is increasing as longer time scales are observed.

We reject objects whose variogram has a computed slope for $h
\geq 100$ days smaller than 0.1.
This limit was established empirically and also studied by doing Monte
Carlo simulations on the SMC data.
If all the signals are composed of pure Gaussian noise, then selecting
the slope higher than 0.1 would select less than 0.1\% of the time
series.  As an other example, if all the signals were periodic with
period of about 10-11 days, the criterion of the slope higher than 0.1
would select less than 0.2\% of the light curves. This threshold will
generate only a small number of false detections.  On the other hand,
a signal with a period between 250 and 300 days would be selected with
this slope criterion with a proportion higher than 99.9\%.

The variable QSO, behind LMC, OGLE050833.29-685427.5 discovered by
Dobrzycki et al. 2002 has its variogram displayed in
Fig.~4. It is clearly selected by our criterion.

By applying the slope selection criterion, the list of QSO candidates
shortens to 649 objects for LMC and 179 for SMC.

\subsection{QSO and Be star colours}

Fig.~5 is a colour-magnitude diagram of the
variable stars and the selected objects in the SMC and LMC.  We see
that there is a very high density of points towards the location of the
main sequence. The magnitude, colour, as well as the light
curve, allow us to identify them with $\gamma\,$Cas stars/Be stars.
For a full discussion of Be stars in SMC, see Mennickent et al.
(2002).  The variation in different bands gives further clues for the
identification of eruptive Be stars: notice that in the brightening
phase the star is becoming brighter in the $I$-band
(cf. Fig.~6).

For the SMC data, we have the data of Zaritsky et al. (2002) which
contain $U$-band photometry. We plot in Fig.~7 the
colour-colour diagram ($U-B$, $B-V$).  We recall that the Johnson
$U$-band is unfortunately covering the Balmer discontinuity at $3647
\mbox{\AA}$ and is not optimal as would be the SDSS bands, for example. This
type of diagram is helpful to identify QSOs (see Fan 1999). Though
QSOs have colours in $B-V$ like RR Lyrae stars, they are brighter in the
$U$-band and can be generally distinguished from stellar locus (more
precisely from A-F stars). There are some other degeneracies,
with white dwarfs, and Be stars for instance.

As we do not have $U$ magnitudes for LMC data, we want to devise a
method applicable to both clouds, which could be checked more
thoroughly with the SMC data.

There is a complication already mentioned above: the Be stars during
their eruption phases become redder and therefore may enter the
domain of the QSOs. A compromise has to be taken to delineate a
reasonable colour cut in $B-V$.

As B stars are generally bluer than QSOs (in $B-V$ colour), we can put
a threshold on $B-V$ colour on the QSO blue side.  We are using the
colours of the QSOs measured by SDSS (Richards et al. 2001), to transform
the $g-r$ colour in $B-V$ using the transformation given by Fukugita
et al. (1996).  For the red side of the Be stars, we use the data of
Mennickent et al. (2002); we selected their Type 1 or Type 2 Be stars.
We clearly want to reject very few QSOs and reject many Be stars.
Putting the cut at $B-V = 0.04$ seems a satisfactory compromise since
it rejects 87\% of Be stars and rejects less that 1\% of the QSOs.

\subsection{Undesirable effects, artifacts and problems}
\label{sec:undeffect}
Some time series are selected due to undesirable effects, some of
those are spurious, others are intrinsic to the star:
\begin{itemize}
  \item Some regions of the CCD chip have a larger number of variables
        than average. We show in Fig.~8 the case of LMC
        and SMC where all variable star positions are plotted in CCD
        coordinate system for the 21 and 11 fields all together
        respectively (see especially for LMC at the left edge of the
        CCD, as well as some horizontal lines. Some features appear
        only in one field; others are present in several fields. The
        field LMC\_SC2 has many perturbations).  In
        Fig.~9, we present the time series of one star
        located in a suspicious region, which has a yearly pattern.  The
        SMC fields are less perturbed.
  \item Brighter or dimmer step-like variation in the data, see.
        Fig.~10.  The identified cause was the
        realuminisation of the mirror which occurred between 19 Jan
        1999 and 22 Jan 1999 (JD: 2451197 - 2451200 days).
  \item Some objects near a bright star can become variable
        because of a bleeding column or extended wings of the
        bright star. We computed the distance from the object to the
        nearest pixel with counts exceeding 30'000 and removed the objects
        with such a distance smaller than 50 pixels.
  \item Small real or spurious monotonic slopes were removed.
  \item Certain objects are appearing several times because of field
        overlap. Those were identified and only one was kept.
  \item The OGLE-II team flagged certain objects as uncertain
        (\.Zebru\'n 2001). Those cases were nearly all removed from our list.
  \item Some stars show periodic variations on short time scale in
        addition to trends at longer time scale. Those stars were
        removed from the list.
\end{itemize}

\section{Results and discussion}

\subsection{SMC data}

The number of objects selected (see section~\ref{sec:magsel}) with the
magnitude and colour cuts was 1553. With the criterion on time scales,
we should expect that the errors of false detection are very few. The
number of selected objects is 179, we introduce the cut on $B-V >
0.04$ which reduces further the number to 45.  This is a small number
of objects and can be easily investigated on individual time series.

There are 15 QSO candidates that are listed in Table~\ref{tab:listSMC}
which contains a total of 25 entries (selected manually). Six objects
are identified as Be stars (identified with colour changes), four
objects are left without identification. We present the light curves
of those objects in Fig.~11.

The object OGLE003850.79-731053.1 is rather bright $I=16.8$. However it was
selected because its $I$-mag mean in the catalogue was of 17.077.

\subsection{LMC data}

The number of objects initially selected for the LMC data is rather
high 6241.  By applying the same time scale criterion as we have for the
SMC, we get 649 objects.  This LMC fraction of 10.4\% is slightly
lower than the SMC fraction of 11.5\%. The stars initially considered
are not completely similar since the SMC is further away while the
magnitude and colour cuts are identical for the two clouds. Therefore
the selection criterion based on time scale rejects the RR Lyrae stars
for the LMC which are not present in the SMC and rejects the short
period Cepheids from the SMC which are too bright to be in the LMC.
However, RR Lyrae stars showing a Blazhko effect are sometimes
selected.

The selection on colour, $B-V > 0.04$ cuts the sample by half to
312 stars. In SMC the sample was cut by one third. This is more
surprising. The population of Be stars in LMC seems therefore to have
a redder extension in $B-V$ colour than the SMC population.

We decided to select manually the 312 stars on individual basis
thereby eliminating the undesirable effects mentioned in
section~\ref{sec:undeffect}. However, we conserved 2 cases classified
as uncertain and also 3 objects with a distance smaller than 50 pixels
to a saturated star.

We end up with a list of 155 candidates see Table~\ref{tab:listLMC}.
and we can see their light curves from Fig.~12 to Fig.~16.  A manual
selection gives 118 QSO candidates, 30 Be stars and 7 unclassified
objects.

\section{Conclusion}

The main result of this study shows that it is possible to establish a
rather narrow list of QSO candidates using mostly photometric times
series in the optical wavelengths. However confirmation with
spectroscopic data is needed. We list 118 and 15 QSO candidates for
the LMC and SMC respectively.

Once the QSOs are eventually identified, we will be able to fine-tune the
algorithm to select QSOs more efficiently.

OGLE-II may give hints of how a sampling could be programmed in order
to optimize the detection of QSOs from a variability point of view.

From the study of Dobrzycki et al. (2002), we remark that to select
QSOs is not a trivial task. From about one hundred candidates, the 30
best objects were selected for spectroscopic follow-up and 4 objects
were confirmed as QSOs.

In a mission like GAIA (Perryman et al. 2001), the number of measured
objects is estimated to be of the order of one billion, the
distinction between QSOs and stars is subject of study with the
current photometric system and astrometric precision (Mignard
2002). Variability could be used as an additional criterion for
selecting QSO candidates and therefore diminishing further the rate of
false detection.

This study is one additional example, that multiepoch surveys like
OGLE, MACHO or EROS originally oriented to detect microlensing events
can be used in many different fields of astronomy. The scientific
outcomes of such surveys are often unexpected.

\section{Acknowledgements}

We would like to thank Prof. B.Paczy\'nski for his help and constant
support.  We are indebted to Prof. M.Grenon, Prof. K.Stanek,
Prof. M.Strauss, Dr M.Freitag, Dr R.H.Lupton and Dr S.Paltani.  We
thank Macauley C.S. Peterson for English corrections.

This work was supported with a grant from the Swiss National Science
Foundation. Partial support for this project was also provided by the
NSF grant AST-9820314.

\onecolumn

\pagebreak
\newpage
\begin{deluxetable}{lcccrrrr}
\tablewidth{11.5cm}
\tabletypesize{\scriptsize} 
\tablecaption{\label{tab:listSMC} The list of selected 25 objects from SMC data.
              It is divided into three parts. QSO candidates QC (15),
              $\gamma\,$Cas/Be stars (6), unclassified U (4). The columns are
              an identification number IN, OGLE identification number, $I$ magnitude,
              number of measurements nmes, V and B magnitudes, the classification,
              and the nearest distance to a saturated pixel}
\tablehead{
\colhead{IN}           &
\colhead{Ident}        &
\colhead{$I$}          &
\colhead{nmes}         &
\colhead{$V$}          &
\colhead{$B$}          &
\colhead{class}        &
\colhead{dist}
}
\startdata
S1  & OGLE003850.79-731053.1 & 17.077 & 293 & 17.694 & 17.741 & QC & 299\\ 
S2  & OGLE004833.68-732955.6 & 18.946 & 298 & 19.459 & 19.584 & QC & 262\\ 
S3  & OGLE004743.68-731630.1 & 17.166 & 298 & 17.520 & 17.652 & QC & 133\\ 
S4  & OGLE004818.25-731242.8 & 17.185 & 297 & 17.443 & 17.490 & QC & 115\\ 
S5  & OGLE004905.88-730257.5 & 17.818 & 311 & 17.964 & 18.038 & QC & 294\\ 
S6  & OGLE005136.59-732016.4 & 17.215 & 303 & 18.018 & 18.400 & QC &  59\\ 
S7  & OGLE005316.80-724219.9 & 18.920 & 307 & 19.270 & 19.528 & QC &  34\\ 
S8  & OGLE005418.94-723737.7 & 17.908 & 306 & 18.163 & 18.211 & QC &  73\\ 
S9  & OGLE005448.97-722544.5 & 18.319 & 244 & 19.017 & 19.179 & QC & 153\\ 
S10 & OGLE005608.34-731911.6 & 19.489 & 272 & 20.036 & 20.280 & QC & 180\\ 
S11 & OGLE010244.86-721521.7 & 18.412 & 278 & 18.890 & 19.392 & QC & 172\\ 
S12 & OGLE010234.69-725424.1 & 17.664 & 278 & 18.372 & 18.699 & QC & 164\\ 
S13 & OGLE010127.63-722422.5 & 18.742 & 279 & 19.067 & 19.211 & QC & 243\\ 
S14 & OGLE010342.76-724419.5 & 18.762 & 270 & 19.471 & 20.156 & QC & 102\\ 
S15 & OGLE010721.61-724845.5 & 18.262 & 268 & 18.958 & 19.207 & QC & 227\\ \hline 
S16 & OGLE003922.09-732531.6 & 17.070 & 292 & 16.926 & 16.850 & Be & 141\\ 
S17 & OGLE003922.07-732531.6 & 17.065 & 266 & 16.942 & 17.029 & Be &  85\\ 
S18 & OGLE004722.13-730844.4 & 17.533 & 299 & 17.688 & 17.728 & Be & 101\\ 
S19 & OGLE004705.03-730611.6 & 17.051 & 299 & 17.325 & 17.382 & Be &  90\\ 
S20 & OGLE004701.81-731650.6 & 17.723 & 299 & 17.861 & 17.987 & Be & 156\\ 
S21 & OGLE005131.37-725054.2 & 17.747 & 311 & 18.075 & 18.141 & Be &  79\\ \hline 
S22 & OGLE004504.35-724449.9 & 17.426 & 241 & 17.948 & 18.438 & U  & 332\\ 
S23 & OGLE004702.90-730800.9 & 17.464 & 299 & 18.057 & 18.566 & U  & 115\\ 
S24 & OGLE005039.13-724154.3 & 18.296 & 303 & 18.852 & 19.138 & U  & 111\\ 
S25 & OGLE005137.19-731429.2 & 17.008 & 290 & 17.255 & 17.298 & U  & 251\\ 
\enddata
\end{deluxetable}

\pagebreak
\newpage
\begin{deluxetable}{lcccrrrr}
\tablewidth{11.5cm}
\tabletypesize{\scriptsize} 
\tablecaption{\label{tab:listLMC} The list of selected 155 objects from LMC
             data. It is divided
             into three parts. QSO candidates (118), Be
             stars (30), unclassified U (7)}
\tablehead{
\colhead{IN}         &
\colhead{Ident}      &
\colhead{I}          &
\colhead{nmes}       &
\colhead{V}          &
\colhead{B}          &
\colhead{class}      &
\colhead{dist}
}
\startdata
L1   &   OGLE05340002-7031278 & 19.008 & 350 & 19.405 & 19.560 & QC &    81 \\ 
L2   &   OGLE05333335-6950083 & 17.939 & 353 & 18.039 & 18.087 & Be &   249 \\ 
L3   &   OGLE05344656-6941542 & 18.200 & 353 & 18.806 & 20.671 & QC &   116 \\ 
L4   &   OGLE05330723-6941091 & 18.126 & 353 & 18.698 & 19.211 & QC &   197 \\ 
L5   &   OGLE05343305-7021375 & 18.252 & 353 & 18.481 & 18.633 & QC &    57 \\ 
L6   &   OGLE05345633-7021385 & 17.030 & 352 & 16.884 & 17.416 & QC &   186 \\ 
L7   &   OGLE05350357-7017506 & 17.200 & 345 & 17.906 & 18.208 & QC &   329 \\ 
L8   &   OGLE05340446-7015224 & 19.076 & 352 & 19.540 & 19.721 & QC &   169 \\ 
L9   &   OGLE05305804-7018345 & 17.534 & 511 & 18.252 & 18.588 & QC &   214 \\ 
L10  &   OGLE05300904-6958289 & 18.525 & 503 & 19.314 & 19.917 & QC &    95 \\ 
L11  &   OGLE05301733-6958358 & 19.450 & 503 & 19.482 & 19.588 & QC &   131 \\ 
L12  &   OGLE05303747-6952233 & 19.413 & 512 & 20.206 & 20.433 & QC &   180 \\ 
L13  &   OGLE05315248-6951473 & 17.648 & 510 & 18.310 & 18.722 & QC &   234 \\ 
L14  &   OGLE05272879-6958124 & 17.962 & 453 & 17.872 & 18.058 & Be &    89 \\ 
L15  &   OGLE05295180-6956585 & 17.359 & 496 & 17.374 & 20.073 & QC &   126 \\ 
L16  &   OGLE05292212-7014254 & 18.372 & 503 & 18.881 & 19.339 & U  &   169 \\ 
L17  &   OGLE05293264-6946295 & 17.853 & 500 & 18.460 & 18.675 & QC &   107 \\ 
L18  &   OGLE05283878-6921034 & 20.303 & 502 & 20.769 & 20.965 & QC &   101 \\ 
L19  &   OGLE05294492-7004584 & 18.037 & 502 & 18.910 & 20.415 & QC &   122 \\ 
L20  &   OGLE05273189-7015093 & 18.327 & 499 & 19.126 & 19.684 & Be &   201 \\ 
L21  &   OGLE05270191-6954425 & 18.520 & 499 & 18.185 & 18.230 & QC &   188 \\ 
L22  &   OGLE05252130-6952397 & 19.440 & 497 & 19.881 & 20.113 & QC &    75 \\ 
L23  &   OGLE05261938-7015347 & 18.257 & 396 & 18.851 & 19.417 & U  &   322 \\ 
L24  &   OGLE05262769-6934247 & 19.192 & 499 & 19.860 & 19.968 & QC &   133 \\ 
L25  &   OGLE05250138-6926069 & 17.493 & 435 & 16.380 & 17.538 & QC &   325 \\ 
L26  &   OGLE05265132-7004130 & 18.099 & 496 & 18.760 & 19.351 & QC &   156 \\ 
L27  &   OGLE05233734-6950300 & 18.732 & 486 & 19.440 & 19.871 & QC &    79 \\ 
L28  &   OGLE05245048-6946337 & 18.869 & 486 & 19.276 & 19.373 & QC &   166 \\ 
L29  &   OGLE05250323-6945242 & 17.589 & 445 & 17.595 & 17.738 & Be &   165 \\ 
L30  &   OGLE05243349-6943234 & 19.195 & 483 & 19.492 & 19.784 & QC &   178 \\ 
L31  &   OGLE05225962-6942158 & 18.041 & 481 & 18.100 & 18.176 & U  &   176 \\ 
L32  &   OGLE05234339-6941370 & 17.619 & 486 & 17.769 & 17.828 & QC &   115 \\ 
L33  &   OGLE05225427-6940093 & 17.108 & 481 & 17.295 & 17.343 & QC &    71 \\ 
L34  &   OGLE05240142-6938349 & 19.605 & 484 & 19.939 & 20.014 & QC &   227 \\ 
L35  &   OGLE05223164-6938197 & 18.119 & 457 & 18.096 & 18.170 & QC &   206 \\ 
L36  &   OGLE05233517-6931597 & 18.196 & 486 & 18.659 & 18.905 & QC &   128 \\ 
L37  &   OGLE05245299-6926264 & 17.561 & 482 & 17.753 & 17.863 & QC &   240 \\ 
L38  &   OGLE05225546-6923456 & 18.473 & 480 & 18.442 & 18.494 & Be &   119 \\ 
L39  &   OGLE05240426-6923450 & 17.701 & 479 & 17.761 & 17.863 & QC &    35 \\ 
L40  &   OGLE05241200-6921181 & 17.870 & 424 & 18.073 & 18.121 & QC &   210 \\ 
L41  &   OGLE05245271-6919417 & 17.453 & 482 & 17.509 & 18.057 & Be &   251 \\ 
L42  &   OGLE05244236-6913582 & 17.996 & 476 & 18.235 & 18.311 & QC &   330 \\ 
L43  &   OGLE05245794-6959169 & 17.522 & 479 & 17.518 & 17.767 & QC &   207 \\ 
L44  &   OGLE05221000-6950209 & 17.295 & 467 & 17.560 & 17.627 & Be &   128 \\ 
L45  &   OGLE05201141-6948494 & 17.043 & 483 & 17.284 & 17.380 & QC &   301 \\ 
L46  &   OGLE05222475-7002083 & 18.077 & 478 & 18.362 & 18.466 & QC &   555 \\ 
L47  &   OGLE05213293-7001010 & 17.129 & 481 & 17.304 & 17.454 & U  &   311 \\ 
L48  &   OGLE05203599-6916251 & 18.049 & 483 & 18.889 & 19.639 & QC &    37 \\ 
L49  &   OGLE05190857-6936529 & 17.287 & 461 & 18.148 & 18.942 & QC &   115 \\ 
L50  &   OGLE05175980-6936072 & 17.583 & 475 & 17.481 & 17.549 & Be &    69 \\ 
L51  &   OGLE05195132-6934301 & 17.144 & 473 & 17.358 & 17.410 & QC &    93 \\ 
L52  &   OGLE05181138-6932336 & 18.378 & 475 & 17.718 & 18.005 & QC &    88 \\ 
L53  &   OGLE05175227-6931485 & 17.860 & 475 & 18.329 & 18.657 & Be &   250 \\ 
L54  &   OGLE05195513-6930534 & 17.136 & 469 & 17.497 & 17.737 & QC &   134 \\ 
L55  &   OGLE05173308-6928039 & 17.514 & 456 & 17.694 & 17.792 & QC &   148 \\ 
L56  &   OGLE05174705-6926485 & 17.784 & 475 & 17.904 & 18.013 & Be &    64 \\ 
L57  &   OGLE05191634-6923391 & 17.483 & 429 & 18.315 & 19.036 & QC &   183 \\ 
L58  &   OGLE05173303-6920189 & 17.371 & 448 & 17.466 & 17.517 & Be &   149 \\ 
L59  &   OGLE05182871-6916480 & 18.702 & 475 & 19.397 & 19.930 & QC &   192 \\ 
L60  &   OGLE05173988-6915438 & 18.208 & 445 & 18.414 & 18.572 & QC &   212 \\ 
L61  &   OGLE05174465-6913307 & 18.336 & 475 & 19.201 & 19.410 & QC &   132 \\ 
L62  &   OGLE05195700-6907017 & 18.246 & 459 & 18.943 & 19.394 & QC &   284 \\ 
L63  &   OGLE05173776-6859282 & 18.064 & 470 & 18.185 & 18.240 & QC &   262 \\ 
L64  &   OGLE05192961-6941229 & 17.566 & 471 & 17.674 & 17.742 & QC &   313 \\ 
L65  &   OGLE05154233-6933390 & 17.364 & 366 & 17.306 & 17.472 & Be &   334 \\ 
L66  &   OGLE05170619-6933232 & 18.105 & 365 & 18.457 & 18.594 & U  &   151 \\ 
L67  &   OGLE05172660-6929392 & 17.646 & 361 & 17.576 & 17.835 & QC &   127 \\ 
L68  &   OGLE05150288-6946549 & 17.954 & 228 & 18.801 & 19.778 & QC &   103 \\ 
L69  &   OGLE05164498-6923022 & 18.181 & 331 & 18.366 & 18.430 & QC &   151 \\ 
L70  &   OGLE05171370-6921071 & 17.976 & 364 & 18.209 & 18.334 & QC &   127 \\ 
L71  &   OGLE05171122-6920326 & 18.879 & 364 & 19.164 & 19.359 & QC &   126 \\ 
L72  &   OGLE05150609-6913527 & 17.781 & 359 & 17.788 & 17.937 & QC &   227 \\ 
L73  &   OGLE05165871-6906427 & 18.781 & 365 & 19.046 & 19.383 & QC &   270 \\ 
L74  &   OGLE05170055-6855280 & 17.755 & 366 & 17.751 & 17.941 & QC &   284 \\ 
L75  &   OGLE05165084-6939052 & 17.340 & 362 & 17.549 & 18.570 & Be &   196 \\ 
L76  &   OGLE05153037-6937072 & 17.432 & 366 & 17.468 & 17.588 & Be &   195 \\ 
L77  &   OGLE05150572-6935308 & 17.738 & 361 & 17.638 & 17.756 & QC &    38 \\ 
L78  &   OGLE05135746-6924434 & 17.923 & 334 & 18.548 & 19.046 & QC &   362 \\ 
L79  &   OGLE05134958-6922569 & 18.352 & 333 & 18.856 & 19.234 & U  &   196 \\ 
L80  &   OGLE05132825-6909559 & 17.795 & 334 & 17.765 & 17.879 & Be &   226 \\ 
L81  &   OGLE05124464-6938573 & 18.653 & 334 & 19.551 & 19.994 & QC &   106 \\ 
L82  &   OGLE05140925-6909202 & 18.749 & 327 & 19.244 & 19.512 & QC &   245 \\ 
L83  &   OGLE05135468-6905033 & 18.698 & 334 & 18.408 & 18.691 & QC &   127 \\ 
L84  &   OGLE05143356-6932456 & 17.744 & 334 & 18.540 & 18.876 & QC &   184 \\ 
L85  &   OGLE05132557-6931217 & 18.792 & 334 & 18.878 & 18.935 & QC &   162 \\ 
L86  &   OGLE05114851-6914204 & 18.186 & 328 & 18.136 & 18.190 & QC &   166 \\ 
L87  &  OGLE05115086-6927307 & 17.203 & 325 & 17.371 & 17.425 & Be &   304 \\ 
L88  &  OGLE05102792-6918227 & 17.516 & 324 & 18.017 & 18.330 & QC &    96 \\ 
L89  &  OGLE05105738-6934202 & 18.108 & 321 & 18.133 & 18.236 & QC &    72 \\ 
L90  &  OGLE05084577-6859573 & 18.484 & 271 & 18.473 & 18.518 & QC &   296 \\ 
L91  &  OGLE05084840-6859339 & 17.537 & 271 & 17.673 & 17.787 & Be &   258 \\ 
L92  &  OGLE05083329-6854275 & 18.625 & 271 & 19.012 & 19.239 & QC &   121 \\ 
L93  &  OGLE05091889-6850316 & 17.494 & 271 & 17.519 & 17.577 & QC &   147 \\ 
L94 &  OGLE05085407-6845091 & 17.368 & 271 & 17.439 & 17.495 & QC &   139 \\ 
L95  &  OGLE05092328-6923182 & 17.691 & 271 & 18.287 & 18.657 & QC &   198 \\ 
L96  &  OGLE05095364-6920549 & 17.200 & 271 & 17.265 & 17.371 & QC &    75 \\ 
L97  &  OGLE05095536-6917217 & 17.602 & 271 & 17.629 & 17.714 & Be &   281 \\ 
L98  &  OGLE05051435-7000510 & 19.373 & 325 & 20.139 & 20.570 & QC &   555 \\ 
L99  &  OGLE05060247-6953112 & 18.140 & 271 & 18.099 & 18.149 & QC &   174 \\ 
L100 &  OGLE05070041-6950078 & 18.864 & 281 & 19.619 & 19.822 & QC &   293 \\ 
L101 &  OGLE05072822-6936252 & 17.608 & 323 & 17.655 & 17.753 & Be &    90 \\ 
L102 &  OGLE05050293-6851166 & 17.209 & 254 & 18.072 & 18.671 & QC &    93 \\ 
L103 &  OGLE05054738-6847018 & 17.109 & 268 & 17.217 & 17.276 & QC &   239 \\ 
L104 &  OGLE05071477-6828385 & 18.709 & 249 & 19.087 & 19.184 & U  &   187 \\ 
L105 &  OGLE05062280-6826175 & 18.375 & 268 & 18.972 & 19.154 & QC &   179 \\ 
L106 &  OGLE05070433-6906561 & 17.304 & 267 & 17.560 & 17.661 & QC &    60 \\ 
L107 &  OGLE05071707-6902003 & 17.178 & 267 & 17.358 & 17.486 & QC &   201 \\ 
L108 &  OGLE05041189-6839319 & 19.120 & 240 & 19.590 & 20.030 & QC &   555 \\ 
L109 &  OGLE05050393-6924088 & 17.484 & 249 & 18.194 & 19.578 & QC &   339 \\ 
L110 &  OGLE05050001-6916150 & 17.458 & 266 & 17.360 & 17.458 & Be &   226 \\ 
L111 &  OGLE05045928-6914385 & 19.831 & 266 & 20.158 & 20.213 & Be &   451 \\ 
L112 &  OGLE05023389-6913062 & 19.326 & 258 & 19.446 & 19.691 & QC &   283 \\ 
L113 &  OGLE05043745-6903563 & 17.317 & 267 & 17.646 & 17.700 & QC &   192 \\ 
L114 &  OGLE05001741-6932160 & 18.208 & 272 & 18.887 & 19.414 & QC &   205 \\ 
L115 &  OGLE05353833-6959577 & 17.534 & 270 & 18.030 & 18.686 & QC &    59 \\ 
L116 &  OGLE05364739-6958500 & 17.172 & 246 & 17.318 & 17.538 & QC &   129 \\ 
L117 &  OGLE05362148-7010259 & 17.286 & 270 & 17.403 & 17.459 & QC &   265 \\ 
L118 &  OGLE05393390-7013031 & 17.260 & 262 & 17.678 & 18.160 & Be &   224 \\ 
L119 &  OGLE05385022-7010177 & 17.148 & 262 & 17.668 & 18.142 & QC &   247 \\ 
L120 &  OGLE05393322-7009017 & 17.202 & 262 & 17.780 & 18.349 & QC &   148 \\ 
L121 &  OGLE05390134-7001496 & 17.731 & 262 & 17.719 & 17.869 & QC &   118 \\ 
L122 &  OGLE05394097-6957016 & 19.296 & 252 & 19.458 & 19.658 & QC &   142 \\ 
L123 &  OGLE05382812-7034559 & 17.649 & 260 & 17.670 & 17.760 & Be &   167 \\ 
L124 &  OGLE05372980-7025049 & 18.330 & 239 & 18.394 & 18.584 & QC &    83 \\ 
L125 &  OGLE05395030-7021529 & 17.520 & 262 & 17.545 & 17.678 & QC &   555 \\ 
L126 &  OGLE05375101-7019354 & 17.153 & 262 & 17.144 & 17.310 & Be &   184 \\ 
L127 &  OGLE05405812-7048157 & 17.109 & 267 & 17.131 & 17.232 & QC &   401 \\ 
L128 &  OGLE05413431-7007009 & 17.585 & 267 & 17.811 & 17.984 & QC &   186 \\ 
L129 &  OGLE05411179-7040148 & 18.006 & 267 & 18.166 & 18.323 & QC &   269 \\ 
L130 &  OGLE05402575-7035356 & 17.772 & 266 & 17.988 & 18.090 & QC &    78 \\ 
L131 &  OGLE05404561-7035579 & 17.701 & 267 & 18.234 & 18.674 & QC &   315 \\ 
L132 &  OGLE05412230-7031251 & 17.816 & 267 & 17.884 & 17.981 & Be &   109 \\ 
L133 &  OGLE05423509-7022233 & 18.118 & 247 & 18.435 & 18.821 & QC &    65 \\ 
L134 &  OGLE05410932-7020155 & 17.031 & 267 & 17.119 & 17.243 & QC &   160 \\ 
L135 &  OGLE05421803-7014363 & 18.128 & 267 & 18.202 & 18.282 & QC &   179 \\ 
L136 &  OGLE05431575-7014130 & 17.391 & 259 & 17.566 & 17.700 & Be &   195 \\ 
L137 &  OGLE05434559-7008569 & 17.109 & 258 & 17.267 & 17.317 & QC &   174 \\ 
L138 &  OGLE05444177-7053214 & 17.448 & 260 & 17.576 & 17.754 & QC &   414 \\ 
L139 &  OGLE05443557-7042037 & 17.563 & 260 & 17.981 & 18.179 & QC &   251 \\ 
L140 &  OGLE05432720-7041332 & 17.170 & 260 & 17.339 & 17.379 & QC &   280 \\ 
L141 &  OGLE05442435-7100170 & 17.113 & 260 & 17.295 & 17.426 & QC &   249 \\ 
L142 &  OGLE05432400-7020232 & 18.319 & 260 & 18.351 & 18.397 & QC &   435 \\ 
L143 &  OGLE05461003-7043283 & 17.778 & 261 & 17.436 & 17.597 & QC &   237 \\ 
L144 &  OGLE05470610-7039006 & 17.115 & 261 & 17.165 & 17.234 & QC &   226 \\ 
L145 &  OGLE05461311-7036397 & 17.170 & 261 & 17.227 & 17.393 & Be &   111 \\ 
L146 &  OGLE05465965-7035430 & 17.581 & 261 & 17.471 & 17.586 & QC &   469 \\ 
L147 &  OGLE05454229-7034300 & 17.346 & 261 & 17.301 & 17.516 & QC &   348 \\ 
L148 &  OGLE05461618-7033250 & 17.712 & 261 & 17.956 & 18.306 & QC &   269 \\ 
L149 &  OGLE05460323-7024215 & 17.500 & 235 & 17.575 & 17.730 & QC &    95 \\ 
L150 &  OGLE05453269-7045239 & 17.643 & 261 & 17.495 & 17.584 & Be &    98 \\ 
L151 &  OGLE05473108-7044592 & 17.564 & 261 & 17.492 & 17.628 & Be &   221 \\ 
L152 &  OGLE05223241-7009470 & 17.721 & 281 & 17.481 & 17.686 & Be &   555 \\ 
L153 &  OGLE05220497-7039356 & 17.878 & 286 & 17.899 & 18.475 & QC &   193 \\ 
L154 &  OGLE05214174-7030291 & 18.578 & 214 & 19.075 & 19.609 & QC &   123 \\ 
L155 &  OGLE05205707-7024528 & 17.494 & 286 & 17.900 & 18.546 & QC &   267 \\ 
\enddata
\end{deluxetable}
\normalsize
\pagebreak
\newpage

\end{document}